\documentclass{article}
\usepackage{spconf,amsmath,graphicx,amssymb, algpseudocode,url}
\usepackage[multiple]{footmisc}
\title{Self-Supervised Anomaly Detection for Narrowband SETI}
\name{Yunfan Gerry Zhang\textsuperscript{1}, Ki Hyun Won\textsuperscript{2}, Seung Woo Son\textsuperscript{3}, Andrew Siemion\textsuperscript{1,4,5,6}, Steve Croft\textsuperscript{1}}
\address{\textsuperscript{1}Department of Astronomy, University of California, Berkeley, CA (yunfanz@berkeley.edu)
\\\textsuperscript{2} Department of EECS, University of California, Berkeley, CA (kihyunwon@berkeley.edu)
\\\textsuperscript{3} Department of EECS, University of California, Berkeley, CA (seungwooson@berkeley.edu)
\\\textsuperscript{4}{SETI Institute, Mountain View, California}
\\\textsuperscript{5}{Radboud University, Nijmegen, Netherlands}
\\\textsuperscript{6}{Institute of Space Sciences and Astronomy, University of Malta}}
\begin{document}
%\ninept
%
\maketitle
\begin{abstract}
The Search for Extra-terrestrial Intelligence (SETI) aims to find technological signals of extra-solar origin. Radio frequency SETI is characterized by large unlabeled datasets and complex interference environment. The infinite possibilities of potential signal types require generalizable signal processing techniques with little human supervision. We present a generative model of self-supervised deep learning that can be used for anomaly detection and spatial filtering. We develop and evaluate our approach on spectrograms containing narrowband signals collected by Breakthrough Listen at the Green Bank telescope. The proposed approach is not meant to replace current narrowband searches but to demonstrate the potential to generalize to other signal types.
\end{abstract}
\section{Introduction}
\label{sec:intro}
\subsection{Overview}
The possibility of detecting radio emission from extraterrestrial technologies was first suggested in 1959 \cite{cocconi}.
%, and was largely motivated by the recognition that early terrestrial communication systems were bright enough to be detectable over large distances (e.g. \cite{klein,cordes,cohen,blair}).
The first modern SETI took place a year later, scanning 400\,kHz of bandwidth in the direction of two stars \cite{drizzydrake}. Such searches for ``technosignatures'' 
%(as a proxy for extraterrestrial intelligence) 
have since greatly expanded in bandwidth, the number of stars surveyed, and the range of signal types detected (e.g. \cite{horowitz,korpela,siemion}).
Launched in 2015, Breakthrough Listen\footnote{\url{http://seti.berkeley.edu/listen}}\footnote{\url{http://www.breakthroughinitiatives.org}} \cite{thegeneral} is the most comprehensive SETI search to date. {\em Listen} is using the Green Bank Telescope (GBT) in West Virginia and the Parkes Telescope in Australia to look at thousands of stars and hundreds of galaxies across multiple GHz of bandwidth. Reduction of raw data results in storage of lower resolution spectrograms at rate of approximately 1PB/year. 
%Data volumes are huge -- even though the raw datasets shrink by a factor 100 when Fourier-transformed into lower-resolution spectrograms, these reduced data products still amount to several PB/year. 

Radio frequency SETI poses considerable signal processing challenge. SETI seeks signals that appear artificial in nature, but that are not originating from earthbound technologies (including Earth-orbiting satellites). Simple signals can be easily detected with a suitable algorithm \cite{enriquez}, but there is a practically infinite parameter space of potential SETI signal types for which hand-coded matched filters may not be effective.
Therefore analysis techniques generalizable to a wider range of signals are necessary. 

The ability to generalize has been one of the main reasons for the success of modern deep learning. Many of the cutting edge deep learning techniques have been developed in the field of computer vision, where large labeled datasets are available. No such comprehensive labeled dataset is currently available for radio astronomy data. Additionally, the morphology of detected signals depends greatly on the frequency band of the receiver, the interference environment of the telescope, and the resolution of the data product, complicating the process of universal labeling. Generative models, which have recently become popular in machine learning, provide alternate approaches to hand-labeling interfering signals. Many forms of generative models are able to learn lower dimensional representations of the data without human supervision. In this work we apply techniques from recent advances in generative deep learning to find signals using anomaly detection. 

\subsection{Problem Formulation}
\label{sec:problem}
The main challenge in radio SETI is contamination from Radio Frequency Interference (RFI) such as that from cellphones, satellites, airplanes or even unintentional emissions such as from microwave ovens. SETI surveys cannot simply reject signals seeminly from human engineered systems, since they may throw out extraterrestrial technosignatures as well. Instead, SETI typically applies spatial filtering to identify signals that appear localized at the target that the telescope is tracking. 
%one (or both) of two criteria: (1) does the signal appear anomalous, unlike any human-generated RFI detected; or (2) does the signal appear to be localized on the sky, i.e. coming from the star or other target that the telescope is tracking. 
{\em Listen} observations at GBT points the telescope at a target star for 5 minutes, then moves off target for 5 minutes, typically iterating this sequence three times per target. If a signal is present in both {\em on} and {\em off}-target observations, it must be from a terrestrial source.
%; if the signal disappears and reappears when the telescope is moved off target and back on again, it is of interest.

In this paper, we attempt to model variations and regularities in the time evolution of signal spectrum from {\em Listen} data to make future predictions. We train a generative model that, when given an {\em on} observation, predicts the {\em off} observation. If an observation fails to match the predictions, an anomaly is triggered. We frame this as a spatial-temporal sequence forecasting problem that can be solved under the general sequence-to-sequence learning framework proposed in \cite{sutskever}. Compared to video sequences in computer vision, radio astronomical data poses different types of challenges. While lacking in the complexities of the high-level features needed for problems such as vehicle detection, astronomical spectrograms typically present high noise environments and signals entangled in interference foregrounds \cite{memisevic}. Besides, the vastly varying strength and spectral spread of signals require a framework highly robust to adversarial perturbations.  

In this paper, we present:

1. An adversarial Convolutional LSTM (ConvLSTM) network that is capable of predicting observations with greatly varying signal strength. 

2. A scheme of anomaly detection based on the predicted and actual observations. 

The rest of this paper is organized as follows. In Section \ref{sec:relatedwork} we discuss some related works of predictive anomaly detection. In Section \ref{sec:models} we discuss the details of our model. In Section \ref{sec:anomaly} we evaluate our model with a scheme of anomaly detection, and in Section \ref{sec:conclusion} we conclude. 

\section{Related Work}
\label{sec:relatedwork}

Recurrent neural network (RNN) and long short-term memory (LSTM) models (\cite{donahue,karpathy,ranzato,srivastava,sutskever,cho,graves}) have driven many of the recent advances in predictive video generation. The LSTM encoder-decoder models proposed in works such as \cite{sutskever} and \cite{srivastava} provide a general framework for sequence-to-sequence learning problems by training two separate LSTM models, one to map input sequence to a vector of fixed dimension and another to extract output sequence from that vector, thereby predicting future sequences or reconstructing past sequences. %There the authors showed good results from the model when applied to moving MNIST digits. 
%In \cite{xingjian}, it is shown that prediction of the precipitation nowcasting can be done by building a single stacked Conv-LSTM in between convolutional and deconvolutional layers used to extract features from input and construct output. 

Predictive anomaly detection in radio frequency data has challenges of high thermal noise and large variations of signal strength. Time domain anomaly detection has first been explored in \cite{oshea_16}. The authors used a recurrent network, and show effective detection of a range of synthetic anomalies. Most recently \cite{tandiya} explores predictive anomaly detection in spectrogram and spectral density functions (SDF). There the authors experimented with simulated signals and anomalies and concluded that there was a greater probability of detection in the SDF domain than in spectrograms.

\section{Method}
\label{sec:models}

\subsection{Dataset}
Breakthrough Listen analysis has largely focused on the search for narrow-band (few Hz bandwidth) signals a few Hz in width\cite{enriquez}. Natural astronomical processes do not produce signals so narrow in frequency. Hence if such a signal can be determined to be not coming from Earth, it is a candidate technological signal from an alien civilization. 
%Another reason for focusing on narrow-band signals is that the argument that such a signal conserve energy, and the fact that such a signal is easy to search with a traditional energy sum pipeline. 
In this paper, we collect the training set with the Doppler-drift narrowband search pipeline \texttt{TurboSETI}, though the technique we propose can be used for more general signal types.

Our dataset was collected from the L-band receiver (1.1 GHz - 1.9 GHz) of the 100 meter Robert C. Byrd Green Bank Telescope in west Virginia, processed into high frequency resolution filterbank files. We extracted 91000 samples that contain narrowband signals. Each sample is cut into an image of size 512 x 16, with 512 frequency stamps and 16 time stamps, and 3Hz by 19sec frequency and time resolutions. We use 90\% of the data for training and the rest for testing.

\subsection{Approach}

We adopt a stacked ConvLSTM model similar to the one described in \cite{srivastava}. Specifically, we adopted composite model, which outputs future prediction as well as input reconstruction. The original motivation for input reconstruction is to help ensure good representation is learned while the predicting decoder tackles prediction. Here we potentially have direct use for the reconstruction decoder in disentangling spatial filtering from actually anomalous signals. For the network to capture context information, we increased the number of input and output frames \footnote{Each frame here is single time stamp of 1D vector.}. Without this we find the model to suffer from the tendency to learn trivial representations that memorizes the last time stamp from the inputs. Operating on eight frames at a time we see qualitative improvements that the model is able to learn patterns in the presence of noise, rather than just producing a constant value. However, the model still shows signs of forgetfulness over longer than eight frames of predictions.

\begin{figure}[htb]

\begin{minipage}[b]{1.0\linewidth}
  \centering
  \centerline{\includegraphics[width=8.5cm]{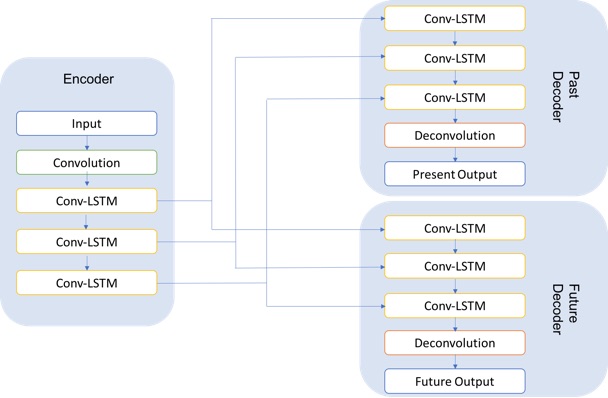}}
%  \vspace{2.0cm}
\end{minipage}
\caption{ConvLSTM Composite Model}
\label{fig:convlstm}
\end{figure}

A key challenge in RF prediction is a loss function that can operate in high noise regimes. Pixel-wise $l_{2}$-loss encourages fitting to not just the signal of interest, but the background noise as well. Adding loss from features extracted with a pre-trained neural network is expected to mitigate this issue. Generative Adversarial Networks \cite{goodfellowgan} has been popular for generating realistic looking images \cite{zhang,wu,ledig,brock}. The idea involves training a generator and discriminator simultaneously with competing goals. The generator is trained to generate samples towards the true data distribution to fool the discriminator, while the discriminator is optimized to distinguish between real samples and fake samples produced by the generator. Further, learning the common behavior of real versus generated signals can be expected to mitigate the issue of network forgetfulness. Thus we solve both of these concerns by introducing a discriminatory network trained concurrently with the ConvLSTM Composite model. The discriminator aims to distinguish between a real and predicted sample, while the generative model incurs an additional $l_{2}$-loss from a high level feature layer of the discriminatory network as well as a distinction loss for the discriminator having successfully identified the predicted sample. 

\begin{figure}[htb]

\begin{minipage}[b]{1.0\linewidth}
  \centering
  \centerline{\includegraphics[width=8.5cm]{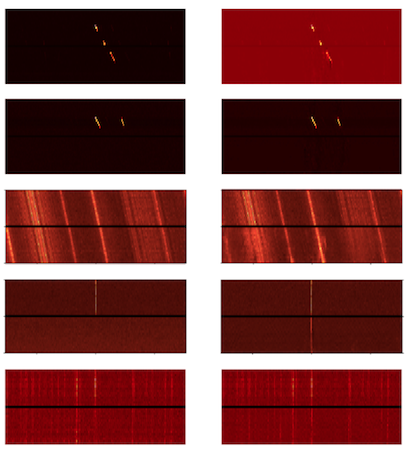}}
%  \vspace{2.0cm}
\end{minipage}
\caption{Example reconstruction and predictions of our network (right) compared to actual observations (left). Each plot shows 10 minutes of observation, with time increasing downward. The line of padding in the middle of each plot shows the moving time of the telescope, when no data is recorded. }
\label{fig:pred}
\end{figure}

We show examples of ground truth as well as reconstructed and predicted on-off pairs in Fig. \ref{fig:pred}. Time increases downward in each plot so that the on-target observation is on top. Left column shows ground truth while right column shows reconstruction and predictions. In the first row the network shows ability to predict intermittent amplitude modulations. The second row shows an example that would have triggered an energy sum detector such as \cite{enriquez}, though our model correctly predicts no signal in the off-frame. The third row shows strongly Doppler-drifting signals. The fourth row shows example of a candidate signal: the network predicts the existence of a signal in the off-frame, while no signal is observed. The last row shows common non-drifting narrowband signals.

\subsection{Implementation Details}
Our Composite model has 8 x 8 filter with 8 number of features for each ConvLSTM layer. Due to the elongated shape of the images, the ConvLSTM layers are sandwiched between one convolutional and one de-convolutional layers with stride 2 in the frequency direction to extract lowest level features without losing crucial information. For experimental training, we feed in the first sequence of 8 frames to predict next 8 frames with batch size of 32. Experiments show that overlapping prediction and input frames leads to marginal improvements. 

We deploy a 5-layer convolutional neural network as the discriminator. Again due to the elongated shape of the images, the discriminator operates on average-pooled input with factor of two reduction in frequency dimension, and a combination of convolutional kernels with asymmetrical strides to reduce the dimension of the image, before a \texttt{tanh} activation unit. All of our models were implemented in \texttt{TensorFlow}.

Our final generative loss function is given by 
\begin{center} 
$\mathnormal{L}$\textsubscript{G} = $\alpha$($\mathnormal{L}$\textsubscript{$\ell$2-future}+$\mathnormal{L}$\textsubscript{$\ell$2-past})+$\beta$$\mathnormal{L}$\textsubscript{$\ell$2-feature}+$\mathnormal{L}$\textsubscript{g}, 
\end{center} 
where $\mathnormal{L}$\textsubscript{$\ell$2-future}+$\mathnormal{L}$\textsubscript{$\ell$2-past} are the pixelwise reconstruction and prediction losses, $\mathnormal{L}$\textsubscript{$\ell$2-feature} is the $\ell$2 loss from features extracted from the 3rd layer of the discriminator, and 
\begin{center} 
$\mathnormal{L}$\textsubscript{g} = $\log$(1-$\mathnormal{D}$($\mathnormal{G}$\textsubscript{future}))\end{center} 
is the usual generative loss for having successfully fooled the discriminator. We take the parameters $\alpha$=0.001 and $\beta$=0.0001. The discriminative loss takes the usual form: 
\begin{center} 
$\mathnormal{L}$\textsubscript{d} = $\log$($\mathnormal{D}$($\mathnormal{G}$\textsubscript{future}))+
$\log$(1-$\mathnormal{D}$($\mathnormal{x}$\textsubscript{future})). 
\end{center}
In the ideal scenario, the generator and discriminator reach  equilibrium and improve simultaneously, in which case $\mathnormal{L}$\textsubscript{g} $\approx$ -0.69, $\mathnormal{L}$\textsubscript{d} $\approx$ -1.38. In practice, however, care must be taken to avoid instability, as we initially observed. To ensure stability, we regularize by monitoring the discriminator and generator loss and update their weights independently to maintain the loss in close range to ideal values. % according to the following algorithm:

% \begin{algorithmic}
% \label{alg:1}
% %\caption{Algorithm to enforce stability of adversarial training. }
% \For{batch in bootstrapped\_dataset}
% \If{$\mathnormal{L}$\textsubscript{g} $>$ -0.5 and $\mathnormal{L}$\textsubscript{d} $>$ -1.0}
%   \State update $\mathnormal{D}$;
%   \State update $\mathnormal{G}$;
% \Else
%   \While{$\mathnormal{L}$\textsubscript{g} $>$ -0.5}
% 	\State update $\mathnormal{G}$ up to 3 times;
%     \EndWhile
  
%   \While{$\mathnormal{L}$\textsubscript{d} $<$ -1.0}
% 	\State update $\mathnormal{D}$ up to 3 times;
%     \EndWhile
%   \EndIf
% \EndFor
% \end{algorithmic} 

\section{Anomaly Detection}
\label{sec:anomaly}
Having trained a predictive model to generate the expected off-target observations, we match it to actual observations to see if anomaly is present. In practice, due to the highly varying signal to noise ratio (SNR), a pixel-wise $l_2$ loss would incur high loss when SNR is high. In \cite{tandiya} the authors split the spectrogram into two dimensional grids and fit a distribution of loss on each grid cell, so that at inference time the log likelihood of the observed loss can be leveraged in predicting anomaly. This ameliorates the effect of noise to a certain extent. However, the authors find the resulting log-likelihood to still vary greatly depending on the signal and hinders accurate anomaly detection. In our case, the very narrow nature of the some of the signals in comparison to our resolution also prohibits gridding. Thus we propose an alternative loss metric for determining anomalies. 

Since we are interested in anomalies that correspond to the spectral-temporal location of the signal, the fluctuations due to varying signal strength is a nuisance. Our new metric is thus centered on the idea to disentangle signal strength from spectral-temporal location. To do this we apply a simple percentile mask to extract the brightest n\% of the pixels from the predicted as well as ground truth observations, and match them by bitwise operations. Specifically, given masks $m_1$ and $m_2$ with filling percentage $n$, we threshold the size of their intersection over the size of their union:
\begin{align*}
	&H_1 \\[-5pt]
	\tau &\gtrless \frac{\|m_1 \& m_2\|}{\|m_1 | m_2\|},\\[-5pt]
	&H_0
\end{align*}
where $\&$ denotes bitwise \texttt{AND} operator and $|$ denotes \texttt{OR}. As sanity check, if the two masks match exactly the above ratio becomes unity, while if they are random and independent, the ratio has expectancy of $n/2\%$. The optimal filling percentage $n$ depends on the band and RFI environment. In practice, we search over a range of thresholds and filling fraction $n$ and find the detection performance to vary little with $1<n<10$. We perform the following experiment to evaluate our method. \footnote{Future plans of this work include a second experiment where we inject and recover signals to evaluate model performance in different regimes of signal to noise ratio. }

\subsection{Pair Matching}
\begin{figure}[htb]

\begin{minipage}[b]{1.0\linewidth}
  \centering
  \centerline{\includegraphics[width=8.5cm]{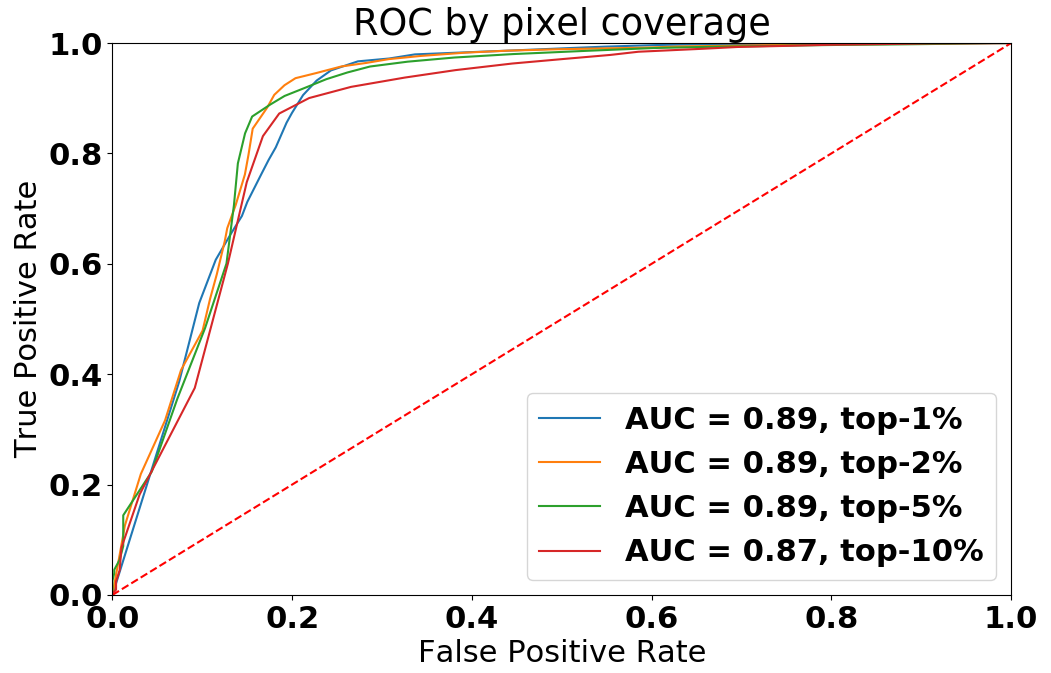}}
%  \vspace{2.0cm}
\end{minipage}
\caption{ROC for the pair matching experiment. The detection performance are roughly self-consistent across filling percentage from 1 to 10. }
\label{fig:ROC_1}
\end{figure}

In this first experiment, we take equal fractions of matching pairs and mismatched pairs of observations, and use our method to identify the two classes. Searching over $\tau$ and $n$ we get the Receiver operating characteristic (ROC) curves shown in Fig. \ref{fig:ROC_1}. The similarity of the curves show that our detection scheme is relatively robust to the choice of $n$. In all cases our schemes are able to correctly identify the pairings with around 90\% AUC. With inspection almost all false alarms correspond to cases where no signal is present in the off-frame observation, where the scheme is thus comparing two noise-like distributions. Combining our scheme with one like \cite{tandiya} to determine if the predicted frame is noise like is expected to further reduce false positive rates in these cases, though doing so is outside the scope of this work. One can also construct the metric for both the reconstruction and future prediction, and comparing the two could help to disentangle spatial filtering from actually anomalous signals.

%\subsection{Signal Injection and Recovery}
%In this second experiment we take matching pairs of observations and for half of them manually inject a narrowband signal in either the ON observation or the OFF observation. The goal is then to retrieve the injected signal. The ROC curve for a range of injected SNR is shown in Fig. With a fixed filling percentage 5\%, 
%

\section{Conclusion}
\label{sec:conclusion}
Radio frequency SETI requires generalizable techniques of data analysis on large unlabeled datasets. In this work, we develop a self-supervised model for radio frequency SETI spatial RFI rejection and anomaly detection. We develop and test our approach on spectrograms of narrowband signals collected by Breakthrough Listen at the Green Bank Telescope. Our model demonstrates ability to learn meaningful predictions of future observations, and even shows signs of improvement over traditional energy sum when the signal appears intermittent. We introduce an anomaly detection scheme that identifies spatially constrained signals in moderate SNR regimes. Our method parallels the efforts of radio frequency anomaly detections in domains of wireless communication. We believe the technique stands ready to be generalized to other signal types, and our successful application to real data in real observational environments encourages fruitful further exploration in this area. 

\vfill\pagebreak

\bibliographystyle{IEEEbib}
\bibliography{ref}

\end{document}